\documentclass[
 reprint, nofootinbib,
 amsmath,amssymb,
 aps,
]{revtex4-2}
 \usepackage[utf8]{inputenc}
\numberwithin{equation}{section}

\usepackage{xcolor}
\usepackage{hyperref}
\hypersetup{
    colorlinks=true,
    citecolor=blue,
    linkcolor=blue,
    filecolor=blue,      
    urlcolor=blue,
    pdftitle={Overleaf Example},
    pdfpagemode=FullScreen,
    }
\usepackage{mathrsfs}
\usepackage{graphicx}
\usepackage{dcolumn}
\usepackage{bm}
\usepackage{orcidlink}
\usepackage{braket}

\begin{document}

\title{Photon Statistics for Fock and Coherent States Interfering in a Beamsplitter}
\author{J. A. T. Santiago}
 \email{student.jhordansantiago@gmail.com}
\affiliation{
Physics Departament, Federal University of Campina Grande, 10071, 58429-900, Campina Grande, Paraíba, Brazil.\\
 }

\date{\today}
\begin{abstract}
We present a straightforward yet comprehensive theoretical study of different quantum states emerging from a bi-modal beamsplitter when various input states interfere. Specifically, we analyze the output states for different combinations of input fields, including Fock states $\ket{n}\ket{m}$, hybrid states $\ket{n}\ket{\alpha}$, and coherent states $\ket{\alpha}\ket{\beta}$. We derive explicit expressions for the output state vectors, calculate the mean photon number, photon number variance, Mandel $Q$ parameter, and second-order coherence function to characterize the statistical properties of the output fields. Our results are intended as a pedagogical resource, serving as an introductory reference for students and researchers aiming to understand basic photon statistics using beamsplitters.

\begin{description}
    \item[Keywords] quantum optics, photon statistics, beamsplitter, q parameter, second-order coherence function.
\end{description}
\end{abstract}

\maketitle
\section{Introduction}

Quantum optics is a field dedicated to studying the quantum nature of light and its interaction with matter \cite{glauber1963coherent, scully1997quantum}. It plays a fundamental role in the development of emerging quantum technologies such as quantum communication \cite{gisin2002quantum}, quantum computation \cite{kok2007linear, ladd2010quantum}, high-precision metrology \cite{giovannetti2004quantum}, and quantum sensing \cite{pirandola2018advances}. The manipulation and control of quantum states of light are essential for these applications, where phenomena like quantum interference, entanglement, and squeezing are routinely exploited \cite{braunstein2005quantum, weedbrook2012gaussian}.

The beamsplitter (BS) is one of the most basic yet powerful devices employed in quantum optics laboratories and experiments \cite{zeilinger1981single}. It enables the coherent superposition of optical modes, leading to interference effects that are central to protocols such as the Hong–Ou–Mandel effect \cite{hong1987measurement} and quantum state tomography \cite{lvovsky2009continuous}. Understanding how various quantum states of light transform through a BS is therefore critical, both for theoretical modeling and experimental implementations.

Over the past decades, numerous experiments have explored the behavior of single-photons, multi-photon Fock states, coherent states, and their superpositions passing through BSs \cite{ou1987violation, rar2005generation,a,b,c}. These studies have revealed rich photon statistics, nonclassical correlations, and phase space interference patterns. Notably, the observation of negative regions in the Wigner function\footnote{The Wigner function provides a representation of a quantum quasiprobability distribution in phase space, a detailed analysis given in \cite{aa,bb}}. serves as a hallmark of nonclassicality and has been confirmed in experiments involving Fock states and Schr{\"o}dinger cat states \cite{deleglise2008reconstruction, ourjoumtsev2007generation}.

A key criterion to characterize the photon number statistics of a quantum state of light is Mandel’s $Q$ parameter \cite{mandel1979sub}, defined by the normalized variance of the photon number distribution relative to a Poisson distribution. 
\textcolor{black}{Such distribution plays a central role in quantum optics as the reference case for photon counting statistics. 
It is a discrete probability distribution describing the likelihood of detecting $n$ photons when the mean photon number is $\langle \mathcal{N} \rangle$, and is written as}
\begin{align}
\textcolor{black}{
P_n= \frac{\langle \mathcal{N} \rangle^n}{n!} e^{-\langle \mathcal{N} \rangle}.}
\end{align}
\textcolor{black}{Its defining feature is that the variance equals the mean, $\Delta \mathcal{N}^2 = \langle \mathcal{N} \rangle$. Distributions broader than Poissonian ($\Delta\mathcal{N}^2 > \langle \mathcal{N} \rangle$) are called super-Poissonian and correspond to photon bunching, where detection events tend to cluster together.  In contrast, narrower distributions ($\Delta \mathcal{N}^2 < \langle \mathcal{N} \rangle$) are called sub-Poissonian and correspond to photon antibunching, that is, the probability of detecting two photons in a very short time interval is smaller than for a Poissonian source, implying that photons tend to arrive one by one rather than in groups, a clear signature of nonclassical light \cite{agarwal2012}}. 

Closely related to $Q$ is the second-order coherence function $g^{(2)}(\tau)$, introduced by Glauber \cite{glauber1963coherent}, which measures temporal correlations between photon detection events. At zero time delay, $g^{(2)}(0)$ can be expressed in terms of the photon number variance and mean, and thus relates directly to Mandel’s $Q$ parameter. \textcolor{black}{Experimentally, $g^{(2)}(\tau)$ is measured using setups such as the Hanbury Brown–Twiss (HBT) interferometer} \cite{hh}. This relation allows one to interpret the photon bunching or antibunching behavior of the field in terms of its number statistics, as we will see later.

In this work, we provide a systematic theoretical analysis of the output quantum states generated by the interference of different classes of input states on a BS. We focus on three representative input configurations: two-mode Fock states $\ket{n}\ket{m}$, hybrid states formed by a Fock state and a coherent state $\ket{n}\ket{\alpha}$, and two-mode coherent states $\ket{\alpha}\ket{\beta}$. For each scenario, we derive explicit analytical expressions for the output states, and compute basic statistical quantities such as the mean photon number, photon number variance, Mandel’s $Q$ parameter, and second-order coherence functions $g^{(2)}(0)$ for the individual output modes. These analyses allow us to investigate how the photon number statistics and nonclassical correlations emerge and evolve as a function of the input state parameters and BS reflectivity/transmissivity.

This work is not intended to provide an exhaustive or highly technical analysis of all possible input-output configurations of BSs. Instead, we focus on fundamental states that allow a clear illustration of photon statistics and the role of quantum versus classical states of light. In this sense, the present discussion is meant to serve as a pedagogical guide, complementing standard quantum optics references and providing an accessible starting point for students and researchers interested in exploring more advanced topics, including quantum correlations and coherence theory.

\subsection{Beamsplitter Theory} 

\begin{figure}[!h]
    \centering
    \includegraphics[width=1\linewidth]{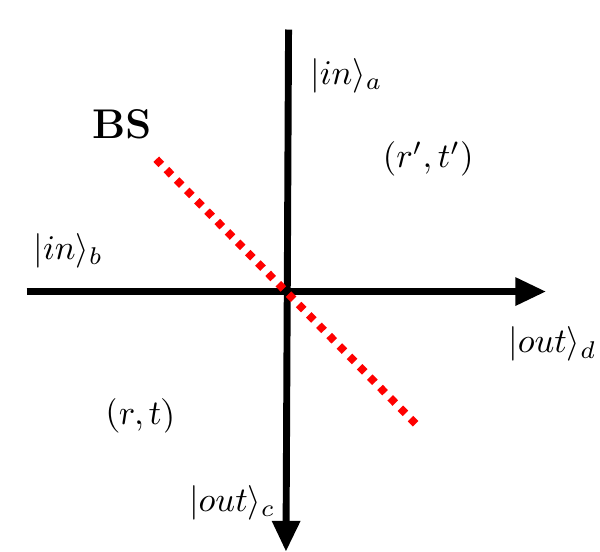}
    \caption{Scheme of a lossless beamsplitter.  
Input modes $a$ and $b$ carry input states $\ket{in}_a$ and $\ket{in}_b$, respectively, 
while output modes $c$ and $d$ are linear combinations of the inputs. 
The BS is characterized by its complex reflection and transmission amplitudes, 
$(r, t)$ for input $a$ and $(r', t')$ for input $b$, which satisfy the unitarity conditions for a lossless device.}
    \label{fig1}
\end{figure}

A bi-modal BS is a fundamental linear optical apparatus that mixes input modes \textcolor{black}{$a$ and $b$} of the electromagnetic field into output modes \textcolor{black}{$c$ and $d$,} through a unitary transformation \cite{gerryknight2005,barnett2002methods}. In the quantum description, the BS acts on the bosonic creation operators $a_a^\dagger$ and $a_b^\dagger$, transforming them into output mode operators $a_c^\dagger$ and $a_d^\dagger$, as seen in Fig. (\ref{fig1}).

A common and widely used model for a lossless BS with complex amplitude reflection coefficient $r$ and transmission coefficient $t$, satisfying \cite{gerryknight2005, xc}
\begin{align}
&|r|=|r'|, \quad |t|=|t'|, \quad |r|^2+|t|^2=1, \\
 &^* t’+ r’ t^* = 0, \quad r^* t+ r’ t’^* = 0,
\end{align}
is given by the transformation:
\begin{equation}
a_c = t a_a + i r a_b, \quad
a_d = i r a_a + t a_b, \label{eq:bs_op_d}
\end{equation}
where the factor $i$ in the reflection terms ensures the overall unitary operation and proper phase relations \cite{Campos1989}.
The inverse transformation reads:
\begin{equation}
a_a = t^*a_c - i r^* a_d,\quad a_b = - i r^*a_c + t^* a_d,\label{x}
\end{equation}
which preserves the bosonic commutation relations:
\begin{equation}
[a_i,a_j^\dagger] = \delta_{ij}, \quad i,j = a,b,c,d.\label{can}
\end{equation}

The unitary operator $U_{BS}$ acts on the two-mode space, such that
\begin{equation*}
a_c = U_{BS}^\dagger a_a U_{BS}, \quad
a_d = U_{BS}^\dagger a_b U_{BS},
\end{equation*}
where
\begin{equation}
U_{BS}(\theta, \phi) = \exp \left[ \theta 
\left( e^{i \phi} a_a^\dagger a_b - e^{-i \phi} a_a a_b^\dagger \right) \right].
\end{equation}

This operator is generated by bilinear combinations of bosonic operators, forming an $SU(2)$ algebra \cite{z,zz,zzz,zzzz}. The Baker--Campbell--Hausdorff (BCH) formula, together with equation \eqref{can}, ensures that the linear transformation in equation~\eqref{eq:bs_op_d} follows consistently from $U_{BS}$ \cite{BachorRalph2003}.

The parameter $\theta \in [0, \pi/2]$ defines the mixing angle, which determines the amplitude transmission and reflection coefficients of the beamsplitter via
\begin{equation*}
|t| = \cos \theta, \quad |r| = \sin \theta.
\end{equation*}
Physically, $\theta$ encodes energy conservation: the sum of reflected and transmitted intensities must equal the input. It also sets how strongly the two modes are mixed: 
\begin{itemize}
    \item $\theta = 0$ corresponds to perfect transmission ($t=1, r=0$), with no mixing of the input modes,
    \item $\theta = \pi/4$ corresponds to a balanced 50:50 beamsplitter ($t = r = 1/\sqrt{2}$), producing maximal superposition of the two input modes,
    \item $\theta = \pi/2$ corresponds to perfect reflection ($t=0, r=1$), effectively swapping the modes.
\end{itemize}

The parameter $\phi$ is, mathematically, up to convention, however, it can also represent a physical phase shift, such as the one accumulated by one of the modes in an interferometric setup. For instance, in a Mach--Zehnder interferometer, a phase delay introduced in one arm appears as $\phi$, leading to observable interference fringes at the output \cite{r,rr}.

For realistic devices, however, losses render the BS transformation effectively non-unitary for the detected modes. These losses can be modeled by coupling each mode to vacuum ancillas, characterized by an overall efficiency $\eta$. The effect of such losses is to reduce the observed mean photon number $\langle \mathcal{N} \rangle$, drive the Mandel $Q$ parameter toward zero, and push the second-order coherence $g^{(2)}(0)$ toward unity. In Hong-Ou-Mandel (HOM) interference, symmetric losses preserve the dip depth but lower overall count rates, whereas asymmetric losses or background noise reduce the observed visibility \cite{lll,l,ll}.
\subsection{Basic Photon Statistics}

The photon number operators for the output modes are:
\begin{equation}
\mathcal{N}_k= a_k^\dagger a_k, \quad k=c, d.
\end{equation}
The mean photon number at each output is \cite{mandelwolf1995}:
\begin{equation}
\langle \mathcal{N}_k \rangle = \mathrm{Tr} \left( \rho \mathcal{N}_k \right).
\end{equation}
Here, $\rho = |out\rangle \langle out|$ is the density operator. The trace $\mathrm{Tr}(\rho \mathcal{O})$ denotes the sum of the diagonal elements of the operator $\mathcal{O}$ in the basis of $\rho$. In other words, it gives the quantum-mechanical average of the observable $\mathcal{O}$ for the system described by $\rho$ \cite{agarwal2012, mandelwolf1995}.

The photon number variance is:
\begin{equation}
(\Delta \mathcal{N}_k)^2 = \langle \mathcal{N}_k^2 \rangle - \langle \mathcal{N}_k \rangle^2\label{variance},
\end{equation}
where $\langle \mathcal{N}_k^2 \rangle = \mathrm{Tr} \left( \rho \mathcal{N}_k^2 \right).$

Mandel's $Q$ parameter, which measures the deviation from Poissonian statistics, is defined as \cite{mandelwolf1995}:
\begin{equation}
Q_k = \frac{(\Delta \mathcal{N}_k)^2 - \langle \mathcal{N}_k \rangle}{\langle \mathcal{N}_k \rangle}.
\end{equation}

In practice, these quantities are computed by expressing $\mathcal{N}_k$ and $\mathcal{N}_k^2$ in terms of the input operators $a_a, a_b$ using the BS transformations, and evaluating their expectation values with the chosen input state.

Another important quantity is the second-order coherence function at zero delay, $g_k^{(2)}(0)$, which characterizes the photon correlation statistics. It is defined as:
\begin{equation}
g^{(2)}_k(0) = 
\frac{\langle \mathcal{N}_k^2 \rangle - \langle \mathcal{N}_k \rangle}{\langle \mathcal{N}_k \rangle^2}.
\end{equation}

The $g^{(2)}(0)$ parameter quantifies the conditional probability of detecting two photons simultaneously at the same output mode, normalized to the square of the mean photon number. It provides further insight into the quantum or classical nature of the light field: $g^{(2)}(0) < 1$ indicates antibunching (nonclassical light), $g^{(2)}(0) = 1$ corresponds to Poissonian (coherent) statistics, and $g^{(2)}(0) > 1$ reveals bunching (typically thermal or chaotic light) \cite{glauber1963coherent}.

Moreover, there is a direct relationship between the Mandel $Q$ parameter and $g^{(2)}(0)$:
\begin{align}
g^{(2)}_k(0) &
=\frac{\langle \mathcal{N}_k^2 \rangle - \langle \mathcal{N}_k \rangle}{\langle \mathcal{N}_k \rangle^2}\nonumber\\
&= 1 + \frac{\Delta \mathcal{N}_k^2 - \langle \mathcal{N}_k \rangle}{\langle \mathcal{N}_k \rangle^2} \nonumber\\
&= 1 + \frac{Q_k}{\langle \mathcal{N}_k \rangle}.\label{g parameter}
\end{align}

\section{Fock State Inputs $|n\rangle |m\rangle$}

A detailed treatment of Fock-state inputs in a beamsplitter can be found in the seminal work of Campos, Saleh, and Teich \cite{Campos1989} and chapter 5 of Agarwal's textbook \cite{agarwal2012}. Following their approach, we consider a two-mode input Fock state of the type
\begin{equation}
|in \rangle = |n\rangle_a |m\rangle_b
= \frac{(a_a^\dagger)^n}{\sqrt{n!}} \frac{(a_b^\dagger)^m}{\sqrt{m!}} |0\rangle_a |0\rangle_b,
\end{equation}
where $|n\rangle_a$ and $|m\rangle_b$ are photon-number eigenstates in the input ports.

Substituting input into output modes through equation \eqref{x}, the state after passing through the BS becomes
\begin{align}
|out \rangle
&= U_{BS} |n\rangle_a |m\rangle_b \nonumber \\
&= \frac{(t a_c^\dagger + i r a_d^\dagger)^n}{\sqrt{n!}}
\frac{(i r a_c^\dagger + t a_d^\dagger)^m}{\sqrt{m!}}
|0\rangle_c |0\rangle_d.\label{xx}
\end{align}

Equation \eqref{xx} describes how a definite number of photons entering the two input ports are redistributed into the output modes $c$ and $d$. Each creation operator in the expansion represents the probability amplitude for a photon to be transmitted or reflected. To obtain the explicit form of the output state in the photon-number basis, we expand each factor using the multi-binomial theorem:
\begin{align}
(t a_c^\dagger + i r a_d^\dagger)^n
&= \sum_{p=0}^{n} \binom{n}{p} t^{n-p} (i r)^{p}
(a_c^\dagger)^{n-p} (a_d^\dagger)^{p}, \\
(i r a_c^\dagger + t a_d^\dagger)^m
&= \sum_{q=0}^{m} \binom{m}{q} (i r)^{m-q} t^{q}
(a_c^\dagger)^{m-q} (a_d^\dagger)^{q}.
\end{align}

Multiplying these expansions, we find
\begin{align}
|out \rangle
&= \frac{1}{\sqrt{n! m!}} \sum_{p=0}^{n} \sum_{q=0}^{m}
\binom{n}{p} \binom{m}{q} t^{n-p+q} (i r)^{m-q+p} \nonumber \\
&\quad \times (a_c^\dagger)^{(n-p)+(m-q)} (a_d^\dagger)^{p+q} |0\rangle_{c} |0\rangle_{d}.
\end{align}

Since
\begin{equation}
(a_c^\dagger)^k (a_d^\dagger)^l |0\rangle_c|0\rangle_{d} = \sqrt{k! l!} |k\rangle_c |l\rangle_d,
\end{equation}
the final expression for the output state is
\begin{align}
|out \rangle
&= \sum_{p=0}^{n} \sum_{q=0}^{m}
\frac{\binom{n}{p} \binom{m}{q}}{\sqrt{n! m!}}
t^{n-p+q} (i r)^{m-q+p} \nonumber \\
&\quad \times \sqrt{(n-p+m-q)! (p+q)!}\nonumber\\
&\quad\times|n-p+m-q\rangle_c |p+q\rangle_d.
\label{eq:output_nm}
\end{align}

Equation \eqref{eq:output_nm} shows that the output is a superposition of two-mode Fock states with all possible photon-number partitions between the two output ports.

\subsection{Mean Photon Number}
Using the BS transformations from equation \eqref{eq:bs_op_d},
we can express the photon number operators at the output as:
\begin{align}
\mathcal{N}_c = a_c^\dagger a_c &= |t|^2 a_a^\dagger a_a + |r|^2 a_b^\dagger a_b + i t r^* a_a^\dagger a_b - i t^* r a_b^\dagger a_a\nonumber \\
&=|t|^2\mathcal{N}_a+|r|^2\mathcal{N}_b+itr^*a_a^\dagger a_b-it^*ra_b^\dagger a_a,\label{28}
\end{align}
\begin{align}
    \mathcal{N}_d = a_d^\dagger a_d &= |r|^2 a_a^\dagger a_a + |t|^2 a_b^\dagger a_b - i t r^* a_a^\dagger a_b + i t^* r a_b^\dagger a_a\nonumber\\
&=|r|^2\mathcal{N}_a+|t|^2\mathcal{N}_b-itr^*a_a^\dagger a_b+it^* ra_b^\dagger a_a\label{29}.
\end{align}

Since the input state is diagonal in the Fock basis, cross terms such as $\langle a_a^\dagger a_b \rangle$ vanish. And, as we know:
\begin{equation*}
\mathcal{N}_\gamma|\gamma\rangle=\gamma|\gamma\rangle,
\end{equation*}
hence, the mean photon numbers are the well-known relations:
\begin{equation}
\langle \mathcal{N}_c \rangle = n |t|^2 + m |r|^2,\quad
\langle \mathcal{N}_d \rangle = n |r|^2 + m |t|^2.\label{mean fock}
\end{equation}

For pure Fock state inputs, these reduce to $\langle \mathcal{N}_c \rangle = n$ and $\langle \mathcal{N}_d \rangle = m$ in the fully transmitting or reflecting limits. However, for mixed states, the mean photon numbers generally differ and must be calculated. Note that the total photon number is conserved:
\begin{equation}
\langle \mathcal{N}_c \rangle + \langle \mathcal{N}_d \rangle = n + m.
\end{equation}
\subsection{Photon Number Variance}

To compute the variance $(\Delta \mathcal{N}_c)^2$, we expand $\mathcal{N}_c^2 = (a_c^\dagger a_c)^2$, with:
\begin{equation}
\langle a_a^\dagger a_a \rangle = n, \quad
\langle (a_a^\dagger a_a)^2 \rangle = n^2,
\end{equation}
\begin{equation}
\langle a_b^\dagger a_b \rangle = m, \quad
\langle (a_b^\dagger a_b)^2 \rangle = m^2,
\end{equation}
therefore, from equation \eqref{28}:
\begin{align}
    \langle\mathcal{N}^2_c\rangle&=\Big\langle\Big(|t|^2n+|r|^2m+itr^*a^\dagger_a a_b-it^*ra^\dagger_b a_a\Big)^2\Big\rangle\nonumber\\
    &=|t|^4n^2+|r|^4m^2+2|t|^2|r|^2nm\nonumber\\
    &\quad+\Big\langle|t|^2|r|^2\Big(a^\dagger_a \underbrace{a_ba^\dagger_b}_{\mathcal{N}_b+1} a_a+a^\dagger_b \underbrace{a_aa^\dagger_a}_{\mathcal{N}_a+1} a_b\Big)\Big\rangle\nonumber\\
    &=\langle \mathcal{N}_c\rangle^2+|t|^2|r|^2\Big((m+1)n+(n+1)m\Big)\nonumber\\
    &=\langle \mathcal{N}_c\rangle^2+|t|^2|r|^2(2nm+n+m).
    \end{align}
The variance is then \cite{Campos1989}:
\begin{equation}
(\Delta \mathcal{N}_c)^2 = |t|^2|r|^2(n+m+2nm).
\end{equation}
\subsection{Mandel's $Q$ Parameter}

The Mandel $Q$ parameter for the $c$-mode is:
\begin{equation}
Q_c = \frac{ (\Delta \mathcal{N}_c)^2 - \langle \mathcal{N}_c \rangle }{\langle \mathcal{N}_c \rangle}.
\end{equation}

Substituting the expressions above:
\begin{equation}
Q_c = \frac{  |t|^2|r|^2(n+m+2nm)}{ n |t|^2 + m |r|^2 }-1, \ \text{for } n=m\neq0.\label{q fock}
\end{equation}

For the vacuum state, $n=m=0$,  the above equation is not directly applicable, as it would involve division by zero. Instead, $Q_c$ for the vacuum can be obtained by continuity, considering it as the $\alpha \to 0$ limit of a coherent state. In this limit, the vacuum is considered to have Poissonian statistics; This is a strictly quantum statement with no classical analogue: the vacuum possesses well-defined quantum fluctuations despite having zero photons on average, illustrating the inherently nonclassical nature of the vacuum \cite{glauber1963coherent}.

\subsection{Second-order coherence function $g^2(0)$}

With equations \eqref{q fock} and \eqref{mean fock} in \eqref{g parameter}, we have for mode $c$:
\begin{align}
    g^2_c(0)=1+\frac{1}{n|t|^2+m|r|^2}\Bigg(\frac{  |t|^2|r|^2(n+m+2nm)}{ n |t|^2 + m |r|^2}-1\Bigg).\label{g fock}
\end{align}

\begin{figure}[!h]
    \centering
    \includegraphics[width=1\linewidth]{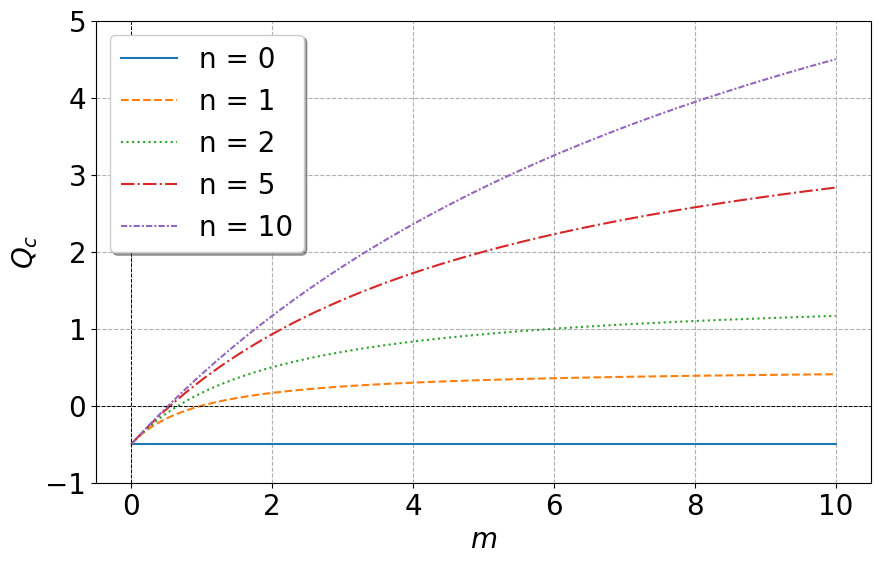}
   \caption{Mandel $Q_c$ parameter for mode $c$ of a 50:50 BS with Fock-Fock state input of the type $|n\rangle_a|m\rangle_b$, where $|r|^2 = |t|^2 = 1/2$. 
    }
        \label{fig q fock}
\end{figure}

\begin{figure}
    \centering
\includegraphics[width=\columnwidth]{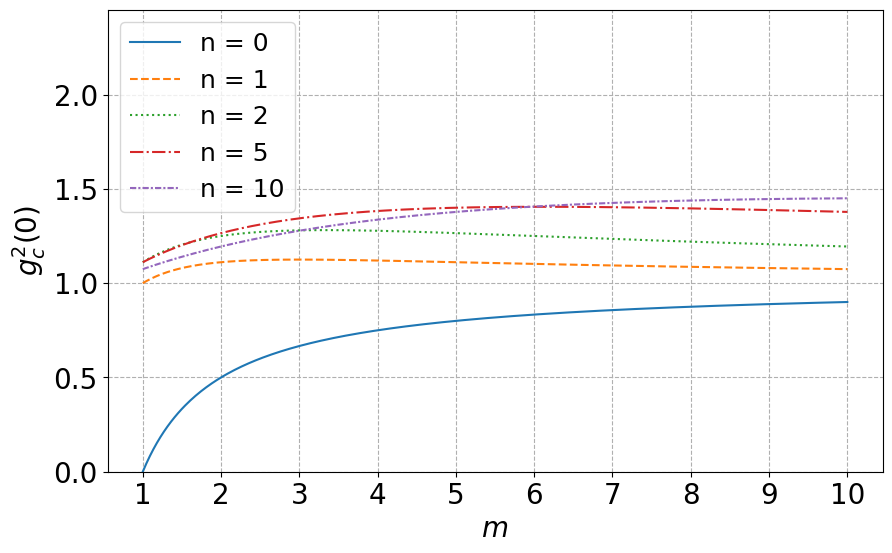}
\caption{$g^{(2)}(0)$ for mode $c$ of a 50:50 BS with Fock-Fock input of the type $|n\rangle_a|m\rangle_b$, where $|r|^2=|t|^2=1/2$.
}
    \label{fig g fock}
\end{figure}

Both parameters from equations \eqref{q fock} and \eqref{g fock} are plotted in figures (\ref{fig q fock}) and (\ref{fig g fock}), respectively. As already mentioned, negative values of $Q_c$ indicate nonclassicality \cite{agarwal2012}, reflecting the sub-Poissonian statistics of the input Fock states, for which the photon-number variance vanishes $(\Delta \mathcal{N})^2 = 0$. However, when $Q_c \geq 0$, no definite conclusion can be drawn. For instance, when $m = n = 1$, the output is a Hong–Ou–Mandel state, yet $Q_c = 0$. The Mandel parameter (and the single-mode $g$ function) captures only local photon-number fluctuations, but does not reveal correlations between modes (which, in the case of the HOM state, are highly quantum). In the limit $m \gg n$, $Q_c$ grows due to the dominance of super-Poissonian statistics. 

This highlights an important distinction: while Mandel's $Q$ and the single-mode $g^{(2)}(0)$ characterize local photon-number fluctuations, they are blind to inter-mode correlations. In practice, the hallmark of the HOM effect is not visible in $Q_c$ or $Q_d$, but rather in the cross-correlation $g^{(2)}_{cd}(0)$, which exhibits a pronounced dip due to the destructive two-photon interference. This shows that full characterization of quantum interference requires both local and nonlocal correlation functions \cite{michek}.

\subsection{Special Cases}

\subsubsection{Single-photon input $|1\rangle_a |0\rangle_b$.}

For a single photon entering the beamsplitter ($n=1$, $m=0$), equation \eqref{eq:output_nm} reduces to
\begin{equation}
|out\rangle_{cd} = t  |1\rangle_{c}|0\rangle_d + i r |0\rangle_{c}|1\rangle_d,
\end{equation}
which represents a photon in a coherent superposition of being in either output mode. It illustrates the fundamental principle of quantum interference at a beamsplitter: even a single photon exhibits wave-like behavior, with probability amplitudes split between the two outputs. Experimentally, this superposition can be observed in single-photon interference experiments using heralded photons from spontaneous parametric down-conversion \cite{brida}.

The photon-number statistics in each output mode are:
\begin{equation*}
\langle \mathcal{N}_c \rangle = |t|^2, \
(\Delta \mathcal{N}_c)^2 = |t|^2|r|^2, 
\ Q_c = - |t|^2, \ g^{(2)}_c(0) = 0,
\end{equation*}
indicating sub-Poissonian statistics and perfect antibunching, characteristic of single-photon states.

\subsubsection{Two photons, one in each input $|1\rangle_a |1\rangle_b$}

For the case $n = m = 1$, the output state is
\begin{equation}
|out\rangle_{cd} =
i r t \sqrt{2}\, |2,0\rangle_{cd} + (t^2 - r^2)\, |1,1\rangle_{cd} + i r t \sqrt{2}\, |0,2\rangle_{cd}.
\label{eq:two_photon_output}
\end{equation}

For a symmetric 50:50 beamsplitter ($|t| = |r| = 1/\sqrt{2}$), the $|1\rangle_c |1\rangle_d$ term vanishes due to quantum destructive interference between indistinguishable photons, giving rise to the  HOM effect \cite{hong1987measurement}, up to a global phase.  Such effect is routinely used to test photon indistinguishability in experiments with single-photon sources, e.g., from quantum dots \cite{ silverstone2016silicon}:

\begin{equation}
|out\rangle_{cd} = \frac{i}{\sqrt{2}} \Big( |2,0\rangle_{cd} + |0,2\rangle_{cd} \Big),
\end{equation}

The photon-number statistics for each output mode are
\begin{align*}
\langle \mathcal{N}_c \rangle = \langle \mathcal{N}_d \rangle = 1, \
(\Delta \mathcal{N}_c)^2 = (\Delta \mathcal{N}_d)^2 &= 1, \\
Q_c = Q_d = 0, \ g^{(2)}_c(0) = g^{(2)}_d(0) = 1,
\end{align*}
Indicating Poissonian statistics for each mode individually, despite the strong quantum correlations between them.
\section{Hybrid Inputs $|n\rangle_a |\alpha\rangle_b$}
We now consider a hybrid state where mode $a$ is a Fock state $|n\rangle$ and mode $b$ a coherent state $|\alpha\rangle$:
\begin{equation}
|in\rangle_{ab} = |n\rangle_a |\alpha\rangle_b
= \frac{(a_a^\dagger)^n}{\sqrt{n!}} |0\rangle_a
\otimes \mathcal{D}_b(\alpha) |0\rangle_b,\label{az}
\end{equation}
where $\mathcal{D}_b(\alpha) = \exp(\alpha a_b^\dagger - \alpha^* a_b)$ is the displacement operator for mode $b$. 
Particularly, the displacement operator shifts the quantum state in phase space: it translates both the quadratures of the field without altering their quantum uncertainties \cite{gerryknight2005}. In this context, a coherent state $|\alpha\rangle$ can be viewed as the vacuum state displaced by $\alpha$ in phase space, i.e., $\mathcal{D}(\alpha)|0\rangle=|\alpha\rangle.$

After the beamsplitter, the output state is
\begin{align}
|out\rangle_{cd}
&= U_{BS} |in\rangle_{ab} \nonumber \\
&= U_{BS} \frac{(a_a^\dagger)^n}{\sqrt{n!}} |0\rangle_a |\alpha\rangle_b \nonumber \\
&= \frac{(t a_c^\dagger + i r a_d^\dagger)^n}{\sqrt{n!}} \mathcal{D}_c(i r \alpha)\, \mathcal{D}_d(t \alpha)\, |0\rangle_c|0\rangle_d,
\end{align}
where we have used the beamsplitter transformations of equation \eqref{x} and the displacement property.

Now, let us expand the operator $(t a_c^\dagger + i r a_d^\dagger)^n$ using the binomial theorem:
\begin{equation}
(t a_c^\dagger + i r a_d^\dagger)^n
= \sum_{k=0}^n \binom{n}{k}\, t^{\,n-k} (i r)^k (a_c^\dagger)^{n-k} (a_d^\dagger)^k.
\end{equation}

Substituting this expansion into the output state, we obtain:
\begin{align}
|out\rangle_{cd} 
&= 
\sum_{k=0}^n \binom{n}{k}\, \frac{t^{\,n-k} (i r)^k}{\sqrt{n!}}  (a_c^\dagger)^{n-k} (a_d^\dagger)^k|ir\alpha\rangle_c|t\alpha\rangle_d.
\end{align}

\subsection{Mean Photon Number}

Using the BS transformations, the mean values for $|n\rangle_a|\alpha\rangle_b$ is:
\begin{equation}
\langle \mathcal{N}_a \rangle =\langle a_a^\dagger a_a \rangle = n, \quad
\langle \mathcal{N}_b \rangle=\langle a_b^\dagger a_b \rangle = |\alpha|^2.
\end{equation}
The mean for the coherent state comes precisely from (one of) its definitions: the coherent state is an eigenstate of the annihilation operator:
\begin{align}
    a|\alpha\rangle=\alpha|\alpha\rangle.
\end{align}
This property implies shot-noise level: relative fluctuations 
scale as $1/\sqrt{\langle \mathcal{N} \rangle}$, and thus coherent states 
define the classical benchmark with $g^{(2)}(0)=1$, separating sub-Poissonian (antibunched) from super-Poissonian (bunched) light.
Hence:
\begin{align}
\langle \mathcal{N}_c \rangle &= n |t|^2 + |r|^2 |\alpha|^2, \quad
\langle \mathcal{N}_d \rangle = n |r|^2 + |t|^2 |\alpha|^2.
\end{align}
Where the total photon number is conserved:
\begin{equation}
\langle \mathcal{N}_c \rangle + \langle \mathcal{N}_d \rangle = n + |\alpha|^2.
\end{equation}
\subsection{Photon Number Variance}
For the variance in mode $c$ $(\Delta \mathcal{N}_c)^2 = \langle \mathcal{N}_c^2 \rangle - \langle \mathcal{N}_c \rangle^2$ , we first calculate $\langle \mathcal{N}_c^2\rangle$
\begin{align}
    \langle\mathcal{N}^2_c\rangle&=\Big\langle\Big(|t|^2n+|r|^2|\alpha|^2+itr^*a^\dagger_a a_b-it^*ra^\dagger_b a_a\Big)^2\Big\rangle\nonumber\\
        &=|t|^4n^2+2|t|^2|r|^2n|\alpha|^2+\langle|r|^4\underbrace{(a^\dagger_b a_b)^2}_{|\alpha|^4+|\alpha|^2}\rangle\nonumber\\
    &\quad+\Big\langle|t|^2|r|^2\Big( a^\dagger_a \underbrace{a_ba^\dagger_b}_{\mathcal{N}_b+1} a_a+a^\dagger_b \underbrace{a_aa^\dagger_a}_{\mathcal{N}_a+1} a_b\Big)\Big\rangle\nonumber\\
    &=\langle \mathcal{N}_c\rangle^2++|r|^4|\alpha|^2+|t|^2|r|^2(2n|\alpha|^2+n+|\alpha|^2),
    \end{align}
hence
\begin{align}
(\Delta \mathcal{N}_c)^2=|r|^4|\alpha|^2+|t|^2|r|^2(2n|\alpha|^2+n+|\alpha|^2).
\end{align}

\subsection{Mandel $Q$ Parameter}

For mode $c$:
\begin{align}
Q_c &= \frac{|r|^4|\alpha|^4+|t|^2 |r|^2 (n + |\alpha|^2 + 2n|\alpha|^2)}{ n |t|^2 + |r|^2 |\alpha|^2 }-1,
  \end{align}

This parameter can be negative (sub-Poissonian) when the Fock state dominates, or positive (super-Poissonian) when the coherent state dominates, reflecting the hybrid quantum-classical nature of the output.
\begin{figure}[!h]
    \centering
    \includegraphics[width=1\linewidth]{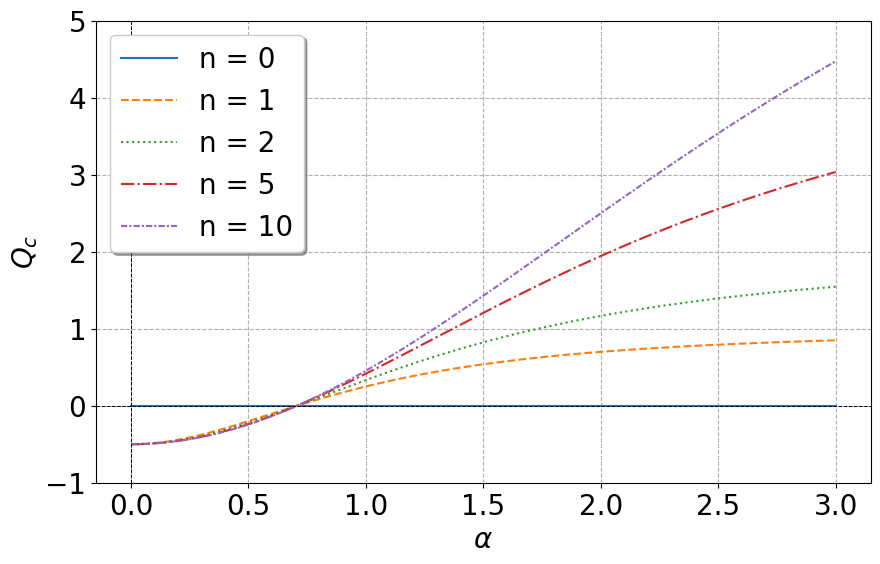}
\caption{
        Mandel $Q_c$ parameter for mode $c$ of a 50:50 BS with Fock-coherent state input of the type $|n\rangle_a|\alpha\rangle_b$, where $|r|^2 = |t|^2 = 1/2$.
        Numerical parameters: $\alpha \in [0,3]$, $n \in \{0,1,2,5,10\}$.
    }
     \label{fig q fock-coherent}
\end{figure}
\begin{figure}[!h]
    \centering
\includegraphics[width=\columnwidth]{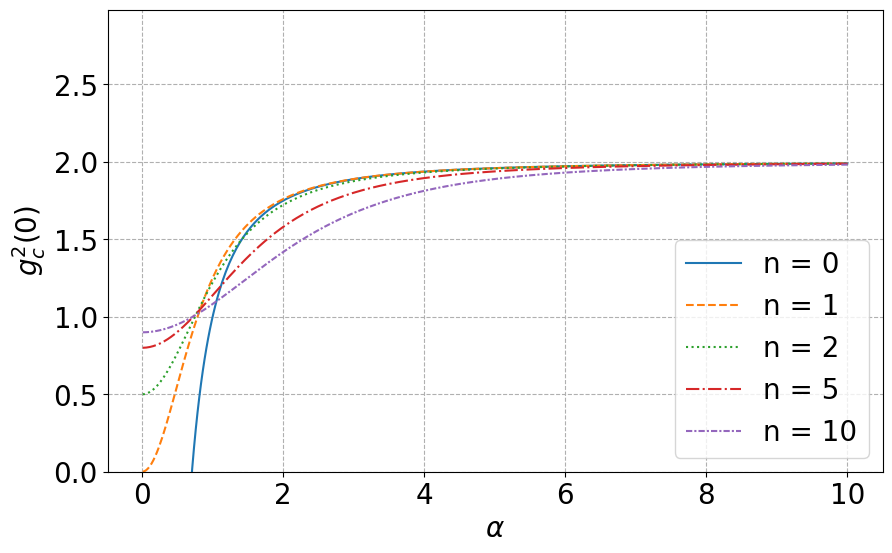}
\caption{$g^{(2)}(0)$ for mode $c$ of a 50:50 BS with Fock-coherent input of the type $|n\rangle_a|\alpha\rangle_b$, where $|r|^2=|t|^2=1/2$.}
    \label{fig g fock-coherent}
\end{figure}
\subsection{Second-order coherence function $g^2(0)$}

For the $c$-mode, the $g$ parameter is:

\begin{align}
    g^2_c(0)&=1+\frac{1}{n|t|^2+|r|^2|\alpha|^2}\nonumber\\
    &\quad\times\Bigg(\frac{|r|^4|\alpha|^4+|t|^2 |r|^2 (n + |\alpha|^2 + 2n|\alpha|^2)}{ n |t|^2 + |r|^2 |\alpha|^2 }-1\Bigg).
\end{align}

$Q$ and $g$ parameters are plotted in figures (\ref{fig q fock-coherent}) and (\ref{fig g fock-coherent}), respectively. For $n=0$ (vacuum in mode $a$), $Q_c=0$ (and $g^2_c(0)=1$), indicating Poissonian statistics. For $n \geq 1$, $Q_c$ becomes negative at small $\alpha$ due to Fock-state antibunching, approaching $Q_c \rightarrow -1/2$ as $\alpha \rightarrow 0$, which reflects the suppression of single-mode photon-number variance. In this regime, particularly for $n=1$ and $\alpha \to 0$, a single photon cannot split into two counts, leading to fluctuations below the Poisson level. As $\alpha$ increases, however, the coherent-state component dominates, introducing phase randomness and driving $Q_c$ toward positive values (super-Poissonian statistics). Equation \eqref{az} thus illustrates the continuous transition between quantum and classical regimes: for small $|\alpha|$, the Fock contribution governs the behavior, yielding antibunching and negative $Q$, whereas for large $|\alpha|$, the coherent field washes out the quantum correlations. This behavior highlights how realistic light sources naturally interpolate between ideal quantum and classical limits.

\subsection{Special Cases}

\subsubsection{Vacuum and coherent state $|0\rangle_a |\alpha\rangle_b$.}
Output state:
\begin{equation}
|out\rangle_{cd} = |i r \alpha\rangle_c |t \alpha\rangle_d.\label{vacuum}
\end{equation}
Statistics remain Poissonian:
\begin{align}
\langle \mathcal{N}_c \rangle =
(\Delta \mathcal{N}_c)^2 = |r|^2 |\alpha|^2, \
Q_c = 0, \ g^2_c(0)=1.
\end{align}
\subsubsection{Single-photon and coherent state $|1\rangle_a |\alpha\rangle_b$.}
Output state:
\begin{equation}
|out\rangle_{cd} = t a_c^\dagger |i r \alpha\rangle_c| t \alpha \rangle_{d} + i r a_d^\dagger |i r \alpha\rangle_c|t  \alpha \rangle_{d}.
\end{equation}
With:
\begin{align*}
\langle \mathcal{N}_c \rangle& = |t|^2 + |r|^2 |\alpha|^2, \\
(\Delta \mathcal{N}_c)^2 &= |r|^4 |\alpha|^2 + |t|^2 |r|^2 (1 + |\alpha|^2 + 2|\alpha|^2),\\
Q_c&=\frac{|\alpha|^4 + 3|\alpha|^2 + 1}{2(1 + |\alpha|^2)} - 1,\\
g^2_c(0)&=1 + \frac{|\alpha|^4 + |\alpha|^2 - 1}{(1 + |\alpha|^2)^2}.
\end{align*}

\section{Coherent State Inputs $|\alpha\rangle_a |\beta\rangle_b$}
We consider the case where both input modes are in coherent states:
\begin{equation}
|in\rangle_{ab} = |\alpha\rangle_a |\beta\rangle_b
= \mathcal{D}_a(\alpha) \mathcal{D}_b(\beta) |0\rangle_a |0\rangle_b,
\end{equation}
where $\mathcal{D}(\xi) = \exp(\xi a^\dagger - \xi^* a)$ is the displacement operator for complex amplitude $\xi$, and $|\alpha\rangle$ and $|\beta\rangle$ are coherent states with amplitudes $\alpha$ and $\beta$, respectively. Coherent states are quantum states that closely resemble classical harmonic oscillators, exhibiting minimum uncertainty and Poissonian photon statistics \cite{glauber1963coherent}. Using the transformation property:
\begin{equation}
U_{BS} \mathcal{D}_a(\alpha) \mathcal{D}_b(\beta) U_{BS}^\dagger = \mathcal{D}_c(t \alpha + i r \beta) \mathcal{D}_d(i r \alpha + t \beta),
\end{equation}
the output state is:
\begin{equation}
|out\rangle_{cd} = |t \alpha + i r \beta \rangle_c |i r \alpha + t \beta \rangle_d.
\label{eq:coherent_output}
\end{equation}

This means that coherent states remain coherent after passing through the BS, with their amplitudes linearly mixed according to the BS coefficients $t$ and $r$.  Importantly, the linearity of the BS ensures that no entanglement is generated between the output modes when both inputs are classical coherent states. This illustrates the close correspondence between quantum coherent states and classical electromagnetic waves, particularly in linear optical systems.

\subsection{Mean Photon Number}

Since coherent states remain coherent under BS transformation:
\begin{align}
\langle \mathcal{N}_c \rangle = |t \alpha + i r \beta|^2, \quad
\langle \mathcal{N}_d \rangle = |i r \alpha + t \beta|^2.
\end{align}

Where the total photon number is conserved:
\begin{equation}
\langle \mathcal{N}_c \rangle + \langle \mathcal{N}_d \rangle = |\alpha|^2 + |\beta|^2.
\end{equation}
This behavior aligns with classical wave theory, where the electric field amplitudes of light waves combine linearly, and the resulting intensity fluctuations follow a Poisson distribution \cite{xc, Campos1989}.
\subsection{Photon Number Variance}
For coherent states, photon number variance equals the mean (since it is Poissonian) \cite{glauber1963coherent, scully1997quantum}:
\begin{align}
(\Delta \mathcal{N}_c)^2 = \langle \mathcal{N}_c \rangle, \quad
(\Delta \mathcal{N}_d)^2 = \langle \mathcal{N}_d \rangle.
\end{align}

\subsection{Mandel Q parameter}

The Mandel $Q$ parameter is zero for both modes:
\begin{align}
Q_c =
Q_d = 0.
\end{align}
This result emphasizes the classical nature of coherent states under linear optics. Since both outputs remain coherent and separable, no entanglement or nonclassical correlations are generated. This property underlies the correspondence between classical interference patterns in optics and their quantum descriptions when restricted to coherent states.

\subsection{Second-order coherence function $g^2(0)$}

Since coherent states follow Poisson statistics, the second-order coherent parameter is 1 for both modes.
\subsection{Special Cases}
\subsubsection{One port in vacuum ($\beta = 0$) $|\alpha\rangle_a|0\rangle_b$}
We recover equation \eqref{vacuum} exchanging $t$ for $r$, given that $|\alpha=0\rangle$ is the same as that of $|n=0\rangle$.

\subsubsection{Balanced inputs and BS ($\alpha = \beta$, $|t|^2 = |r|^2 = 1/2$).}
Output amplitudes:
\begin{align}
t \alpha + i r \beta &= \frac{1}{\sqrt{2}} (\alpha + i\alpha) = \frac{(1+i)\alpha}{\sqrt{2}}, \nonumber\\
i r \alpha + t \beta &= \frac{1}{\sqrt{2}} (i\alpha + \alpha) = \frac{(1+i)\alpha}{\sqrt{2}}.
\end{align}
Thus:
\begin{equation}
|out\rangle_{cd} = \left| \frac{(1+i)\alpha}{\sqrt{2}} \right\rangle_c \left| \frac{(1+i)\alpha}{\sqrt{2}} \right\rangle_d.
\end{equation}
Each output mode has mean photon number $|\alpha|^2$.
\section{Conclusions}

In this work, we have presented a theoretical analysis of the photon statistics resulting from the interference of quantum light fields at an ideal, lossless beamsplitter (BS). 
By exploring a broad range of input states, namely Fock--Fock, Fock--coherent, and coherent--coherent configurations, we derived explicit analytical expressions for the output states and computed fundamental statistical quantities, including the mean photon number, photon number variance, Mandel $Q$ parameter, and $g$ function. These calculations were carried out using standard operator techniques and the well-known properties of coherent and Fock states, which highlight the quantum structure underlying the field transformations induced by the BS.

Our results show the close interplay between interference and quantum statistics in shaping the output photon distributions. In particular, for specific Fock state inputs, we recovered nonclassical features such as sub-Poissonian statistics (e.g., when sending vacuum and single-photon states in the input modes) and the Hong--Ou--Mandel interference effect (for a single-photon ejected at each input mode of a balanced BS), which originates from the indistinguishability and bosonic nature of photons. For hybrid inputs combining Fock and coherent states, we analyzed how the coherent component partially preserves its classical-like statistical behavior while still exhibiting interference effects due to the quantum nature of the Fock component. In the case of purely coherent inputs, we confirmed that the output states remain coherent and retain their Poissonian photon number distributions, with the interference being fully determined by the adjustable BS coefficients $t$ and $r$.

In addition to these statistical moments, we also computed the normalized second-order correlation function at zero delay, $g^{(2)}(0)$, for each output mode. This quantity provides a deeper insight into the photon bunching and antibunching behavior of the output fields, revealing regimes of classical and nonclassical light. For Fock--Fock inputs, we recovered exact expressions demonstrating antibunching effects for single-photon inputs and bunching for higher photon numbers. In the case of Fock--coherent interference, our results captured how the presence of a coherent component smooths out the nonclassical correlations, interpolating between sub-Poissonian and Poissonian statistics depending on the relative intensities. These analytical expressions for $g^{(2)}(0)$ help elucidate the transition from quantum to classical behavior in beamsplitter outputs, and offer practical utility in characterizing sources and designing experiments in quantum optics.

Beyond these specific cases, our study provides a unified perspective on how different statistical regimes, ranging from purely quantum (sub-Poissonian) to classical-like (Poissonian), emerge from the linear mixing of quantum states at a BS. This viewpoint is particularly relevant for understanding fundamental experiments in quantum optics, such as single-photon interference and coherent state splitting, as well as for interpreting modern applications in quantum information processing and photonic technologies. However, we note that the model considered here is intentionally simplified and didactic. In practical implementations, additional degrees of freedom, especially photon polarization and the spectral (frequency) dependence of the field operators, $a = a(\omega)$, play a crucial role in determining the observed interference and photon statistics. For a detailed treatment of these effects in multi-photon interference and their application to measurement-device-independent quantum key distribution, see \cite{padrana2019}.

Finally, by presenting the derivations and results in a clear and accessible way, we aim to provide a complementary reference and an intuitive framework for students and researchers interested in the quantum statistical properties of light in a beamsplitter. We hope that this study can serve as a starting point for more advanced investigations of quantum interference phenomena and their applications.

\begin{acknowledgments}
The author gratefully acknowledges the encouragement of Prof. Danieverton Moretti, the constructive criticisms and suggestions of the anonymous referees, and the financial support provided by UFCG.
\end{acknowledgments}

\end{document}